\documentclass[onecolumn,amsmath,amssymb,12pt,superscriptaddress,nofootinbib]{revtex4-2}
\pdfoutput=1

\usepackage[english]{babel}
\usepackage{amsthm}
\usepackage{amssymb}
\usepackage{amsmath}
\usepackage{amsthm}
\usepackage[]{graphicx}
\usepackage{tensor}
\usepackage{color}
\usepackage{framed}
\usepackage{framed}
\usepackage[bookmarks,linktocpage, colorlinks=true, plainpages = false, citecolor = blue,  linkcolor=blue, urlcolor = blue, filecolor = blue]{hyperref} 
\usepackage{natbib}
\usepackage{dcolumn}
\usepackage{enumerate}
\usepackage{bm}
\usepackage{graphicx}
\usepackage{caption}
\usepackage{subcaption}
\graphicspath{{figures/}}

\theoremstyle{definition}

\usepackage{tikz}
\usetikzlibrary{decorations.pathmorphing, patterns,shapes}

\input epsf

\begin{document}
\allowdisplaybreaks
\title{Complex evaluation of angular power spectra:\protect\\ Going beyond the Limber approximation}
\author{Job Feldbrugge}
\email{job.feldbrugge@ed.ac.uk}
\affiliation{Higgs Centre for Theoretical Physics, University of Edinburgh, 
James Clerk Maxwell Building, Edinburgh EH9 3FD, UK}

\begin{abstract}
Angular power spectra are central to the study of our Universe. In this paper, I develop a new method for the numeric evaluation and analytic estimation of the angular cross-power spectrum of two random fields using complex analysis and Picard-Lefschetz theory. The proposed continuous deformation of the integration domain resums the highly oscillatory integral into a convex integral whose integrand decays exponentially. This deformed integral can be quickly evaluated with conventional integration techniques. These methods can be used to quickly evaluate and estimate the angular power spectrum from the three-dimensional power spectrum for all angles (or multipole moments). This method is especially useful for narrow redshift bins, or samples with small redshift overlap, for which the Limber approximation has a large error.
\end{abstract}

\maketitle

\section{Introduction}
In cosmology, we often observe realizations of random fields. The quantum fluctuations of the early Universe lead to a specific realization of radiation and matter in the cosmic microwave radiation field and the present-day cosmic web. Consequently, cosmological surveys often focus on the correlations of these random fields and the cross-correlation functions between them. Consider for example the cosmic microwave background anisotropies, the fluctuations in the density and galaxy distribution, the weak lensing shear and convergence fields, and 21cm emission line fluctuations. In cosmology, these random fields are often characterized by $N$-point correlation functions or by their Fourier transforms known as the power spectra. These fluctuation spectra are important as they are used to test the physics of the early Universe, its contents, and the nature of gravity. However, many of these observations are measured in terms of an angular correlation function $w_{AB}(\hat{\bm{n}}\cdot \hat{\bm{m}})$ of fields $A$ and $B$ and the points on the celestial sphere $\hat{\bm{n}}, \hat{\bm{m}}$, as it is often easier to measure angular positions than proper distances. Instead of working with the angular two-point correlation function directly, we often consider the spherical harmonic transform $C_{AB}(l)$, defined as
\begin{align}
  w_{AB}(\hat{\bm{n}}\cdot \hat{\bm{m}}) 
  = \sum_{l=0}^\infty \frac{2l+1}{4\pi} C_{AB}(l) P_l(\hat{\bm{n}}\cdot \hat{\bm{m}})\,,
\end{align}
with the Legendre polynomial $P_l$. Predictions of the angular power spectra often involve the projection of three-dimensional power spectra onto the celestial sphere, requiring the numerical evaluation of multi-dimensional oscillatory integrals, in particular involving the spherical Bessel transformation of radial selection kernels. These transformations are unfortunately generally expensive to evaluate using standard numerical methods. 

The Limber approximation \cite{Limber:1953} and its generalization to Fourier space \cite{Kaiser:1992, Kaiser:1998} are common methods to approximate the spherical Bessel transform and estimate the projection. More recently, the Limber approximation was extended to higher orders \cite{LoVerde:2008}. These approximations assume small angular separations (or large multipole moments $l$) and that the functions being integrated are slowly varying. The Limber approximation and its extensions are powerful methods, that accurately estimate the magnitude, lead to an analytic understanding of how the projected power spectra depend on the projection procedure and greatly simplify its evaluation. For an analysis of the Limber approximation and proposed alternative approximations for the real space correlation function see \cite{Simon:2007, Schmidt:2008, Lemos:2017}. 

However, these approximations do not always suffice, in particular when considering large angular separations (low multipole moments $l$) and quickly varying radial selection kernels. Given the present and next generation of cosmological surveys, it is becoming increasingly important to quickly project three-dimensional power spectra to angular power spectra and go beyond the Limber approximation. There indeed exists a rich literature on numerical methods, approximating the spherical Bessel transform \cite{Siegman:1977, Talman:1978, Sharafeddin:1992, Lemoine:1994, Talman:2009, Toyoda:2010} and the corresponding projection to angular power spectra \cite{Assassi:2017, Campagne:2017, Schoneberg:2018, Fang:2020, Bella:2021}. Recently, some of these methods were compared in preparation for the Legacy Survey of Space and Time (LSST) \cite{Leonard:2023}.
 
In this paper, I use Picard-Lefschetz theory, an application of Cauchy's integral theorem in complex analysis, to propose a new and efficient method to rephrase the spherical Bessel transform of the Gaussian kernel into a convex integral with no oscillations. This method is subsequently extended to the spherical Bessel transform of the linear combinations of Gaussian kernels spanning a large class of functions. The proposed method is simpler than the previously proposed numerical schemes. Moreover, it leads to a saddle point approximation of the spherical Bessel transform that is accurate in a large range of parameter space, complementing the traditional Limber approximation and its extensions. Note in particular that the saddle point scheme becomes increasingly accurate for narrow selection kernels, where the Limber approximation and its extensions fail.

In section \ref{sec:Projection}, I derive the projection equation and define the notation used in this paper. In section \ref{sec:Limber}, I briefly summarize the Limber approximation and its extension. Section \ref{sec:PicardLefschetz} contains the central results of this paper. I summarize the key points of Picard-Lefschetz theory and illustrate how the technique can be used to reformulate the spherical Bessel transform into an integral without oscillations. Moreover, I present a saddle point approximation that can be used to estimate both the spherical Bessel transformation and the angular power spectrum. Section \ref{sec:Comparison} compares the proposed integration method with the results obtained with a brute force evaluation. I subsequently demonstrate the use of the saddle point approximation and compare it with the Limber approximation and its extension. Concluding remarks are given in section \ref{sec:Conclusions}.

\section{Projection onto the sky}\label{sec:Projection}
Let's consider two random fields $A(\bm{x})$ and $B(\bm{x})$ with their Fourier transforms 
\begin{align}
  \hat{A}(\bm{k}) = \int_{\mathbb{R}^3} A(\bm{x}) e^{-i\bm{k}\cdot \bm{x}}\mathrm{d}\bm{x}\,,\quad
  \hat{B}(\bm{k}) = \int_{\mathbb{R}^3} B(\bm{x}) e^{-i\bm{k}\cdot \bm{x}}\mathrm{d}\bm{x}\,.
\end{align}
The random fields can represent the density fluctuations $\delta \rho(\bm{x})$, the temperature fluctuations $\delta T$, or the Newtonian gravitational potential $\Phi(\bm{x})$. The cross-correlation power spectrum $P_{AB}(k)$ of statically homogeneous and isotropic fields $A$ and $B$ is defined as
\begin{align}
  \langle \hat{A}(\bm{k}_1) \hat{B}^*(\bm{k}_2) \rangle = (2\pi)^3 \delta^{(3)}(\bm{k}_1-\bm{k}_2)P_{AB}(k_1)\,,
\end{align}
with the norm $k_1=\|\bm{k}_1\|$ and the three-dimensional Dirac delta function $\delta^{(3)}$. To evaluate the angular power spectrum, I project these random fields onto the sky with the projection kernels $F_A(r)$ and $F_B(r)$ representing the sensitivity of the survey in the radial direction
\begin{align}
  \tilde{A}(\hat{\bm{n}}) =  \int_0^\infty F_A(r) A(r \hat{\bm{n}})\mathrm{d}r\,,\quad
  \tilde{B}(\hat{\bm{n}}) =  \int_0^\infty F_B(r) B(r \hat{\bm{n}})\mathrm{d}r\,.
\end{align}
Expand $\tilde{A}$ and $\tilde{B}$,
\begin{align}
  \tilde{A}(\hat{\bm{n}}) = \sum_{l=0}^\infty \sum_{m=-l}^l A_{lm} Y_{lm}(\hat{\bm{n}})\,,\quad
  \tilde{B}(\hat{\bm{n}}) = \sum_{l=0}^\infty \sum_{m=-l}^l B_{lm} Y_{lm}(\hat{\bm{n}})\,,
\end{align}
in terms of spherical harmonics $Y_{lm}(\hat{\bm{n}})$ with the harmonic coefficients
\begin{align}
  A_{lm} = \int_{\mathbb{S}^2} \tilde{A}(\hat{\bm{n}})Y_{lm}^*(\hat{\bm{n}}) \mathrm{d}\hat{\bm{n}}\,,\quad
  B_{lm} = \int_{\mathbb{S}^2} \tilde{B}(\hat{\bm{n}})Y_{lm}^*(\hat{\bm{n}}) \mathrm{d}\hat{\bm{n}}\,.
\end{align}
Using the Rayleigh plane-wave expansion
\begin{align}
  e^{i\bm{k}\cdot \bm{x}} = 4 \pi \sum_{l=0}^\infty \sum_{m=-l}^l i^l j_l(k r) Y_{lm}^*(\hat{\bm{k}}) Y_{lm}(\hat{\bm{n}})\,,
\end{align}
these harmonic coefficients can be expressed in terms of the Fourier transform of the random fields
\begin{align}
  A_{lm} &= \frac{i^l}{2\pi^2} \int_{\mathbb{R}^3} A(\bm{k})Y_{lm}^*(\hat{\bm{k}}) \left[\int_0^\infty F_A(r)  j_l(kr) \mathrm{d}r \right] \mathrm{d}\bm{k}\,,\\
  B_{lm} &= \frac{i^l}{2\pi^2} \int_{\mathbb{R}^3} B(\bm{k})Y_{lm}^*(\hat{\bm{k}}) \left[\int_0^\infty F_B(r)  j_l(kr) \mathrm{d}r \right] \mathrm{d}\bm{k}\,,
\end{align}
where $\bm{k} = k \hat{\bm{k}}$ with the norm $k=\|\bm{k}\|$ and the angular position $\hat{\bm{k}}$ with $\|\hat{\bm{k}}\|=1$.

\begin{figure}
  \centering
  \includegraphics[width=0.5\textwidth]{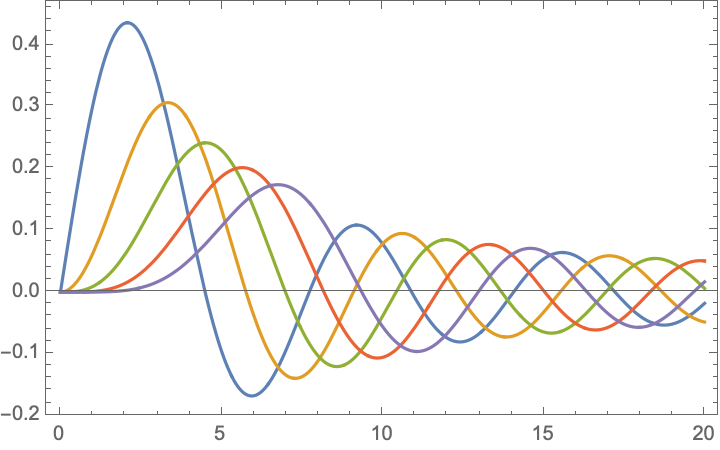}
  \caption{The spherical Bessel function $j_{l}(x)$ for $l=1,\dots,5$ respectively in blue, yellow, green, red, and purple.}
  \label{fig:Bessel}
\end{figure}

The angular power spectrum, projecting the three-dimensional power spectrum onto the celestial sphere, is defined as the three-dimensional oscillatory integral
\begin{align}
  C_{AB}(l) 
  &= \langle A_{lm} B_{lm}^*\rangle\\
  &= \frac{1}{4 \pi^4} \iint_{\mathbb{R}^3\times\mathbb{R}^3} \langle \hat{A}(\bm{k})\hat{B}^*(\bm{k}')\rangle Y_{lm}^*(\hat{\bm{k}}) Y_{lm}(\hat{\bm{k}}')\nonumber\\
  &\phantom{= \frac{1}{4\pi^4 \iint}} \times \left[ \int_0^\infty F_A(r_1) j_l(k r_1)\mathrm{d}r_1\right] \left[\int_0^\infty F_B(r_2)j_l(k r_2)\mathrm{d}r_2\right] \mathrm{d}\bm{k}\mathrm{d}\bm{k}'\\
  &= \int_0^\infty  \frac{2k^2 P_{AB}(k)}{\pi} \left[ \int_0^\infty F_A(r_1) j_l(k r_1)\mathrm{d}r_1\right] \left[\int_0^\infty F_B(r_2)j_l(k r_2)\mathrm{d}r_2\right]
     \mathrm{d}k\\
  &= \int_0^\infty  \frac{2k^2 P_{AB}(k)}{\pi} \mathcal{F}^A_{l}(k)\mathcal{F}^B_{l}(k)
  \mathrm{d}k\,. \label{eq:projection}
\end{align}
The integral over the Fourier mode $k$ is well-behaved as the power spectrum $P_{AB}$ generally decays for small and large Fourier modes. On the other hand, the (modified) spherical Bessel transform of the radial kernels $F_A$ and $F_B$,
\begin{align}
  \mathcal{F}_l(k) = \int_0^\infty F(r) j_l(k r)\mathrm{d}r\,,
  \label{eq:SBT}
\end{align}
converges due to the cancelation of many oscillations (see fig.\ \ref{fig:Bessel} for the first few spherical Bessel functions of the first kind). This integral is generally expensive to evaluate for large multipole moments $l$ along the real line\footnote{Traditionally, the spherical Bessel transform is defined with an additional $r^2$ term, \textit{i.e.}, $\int_0^\infty F(r) j_l(kr) r^2\mathrm{d}r$. We can always transform between the two definitions with a redefinition of the kernel $F \mapsto r^2 F$.}. From hereon, I drop the labels $A$ and $B$.

\section{Limber approximation and beyond}\label{sec:Limber}
The Limber approximation is built on the intuition that the first peak of the spherical Bessel function, at approximately $k r=l+1/2$, dominates in the spherical Bessel transform \cite{Limber:1953}. The subsequent oscillations cancel leading to an insignificant contribution. Formally, for large multipole moments $l$, the spherical Bessel function is replaced by a Dirac delta function $\delta^{(1)}$ centered at the first peak,
\begin{align}
  j_l(x) \mapsto \sqrt{\frac{\pi}{2l+1}}\delta^{(1)}(l+1/2-x)\,,
\end{align}
yielding the simple result
\begin{align}
  \mathcal{F}_l(k) \approx \sqrt{\frac{\pi}{2l+1}} \frac{1}{k}F\left(\frac{2l+1}{2k}\right)\,.
\end{align}
The angular power spectrum reduces to a one-dimensional integral over the Fourier mode,
\begin{align}
  C_{AB}(l) 
  &\approx \int_0^\infty  \frac{2 P_{AB}(k)}{2l+1} F_A\left(\frac{2l+1}{2k}\right) F_B\left(\frac{2l+1}{2k}\right)
     \mathrm{d}k\,,
\end{align}
which is evaluated with standard numerical methods, such as the Gaussian quadrature scheme.

Recently, the Limber approximation was extended \cite{LoVerde:2008} to include higher-order derivatives of the selection kernel using the expansion
\begin{align}
  \mathcal{F}_l(k) \approx \sqrt{\frac{\pi}{2k}} \left[ 
    k^{-1}f\left(\frac{2l+1}{2k}\right) 
    + 
    \frac{k^{-3}}{2}f''\left(\frac{2l+1}{2k}\right) 
    -
    \frac{k^{-4}(2l+1)}{12}f'''\left(\frac{2l+1}{2k}\right) 
    +\dots
    \right]\,,
\end{align}
where the kernel $f$ is defined as $f(r)=F(r)/\sqrt{r}$. After some manipulation, this yields the second-order approximation
\begin{align}
  C_{AB}(l) &\approx  \int_0^\infty \frac{P_{AB}(k) f_A(\frac{2l+1}{2k})f_B(\frac{2l+1}{2k})}{k}\nonumber\\
  &\phantom{\approx \int_0^\infty} \times \bigg[1+ \frac{2}{(2l+1)^2} \left[
    \frac{\mathrm{d} \ln f_A}{\mathrm{d}\ln r}
    \frac{\mathrm{d} \ln f_B}{\mathrm{d}\ln r} s(k) - p(k)
  \right] + \mathcal{O}\left((l+1/2)^{-4}\right)
  \bigg]\mathrm{d}k\,,
\end{align}
with the auxiliary functions
\begin{align}
  s(k)=\frac{\mathrm{d}\ln P_{AB}(k)}{\mathrm{d} \ln k}\,,
  \quad
  p(k) = \frac{k^2(3 P_{AB}''(k) + k P_{AB}'''(k))}{3 P_{AB}(k)}\,.
\end{align}
This approximation is accurate to second-order in $(l+1/2)^{-1}$. Note that these approximations work well for large multipole $l$ and slowly varying kernels $F_A$ and $F_B$.

More recently, new numerical methods were developed to evaluate the projection integral \eqref{eq:projection}. These methods range from smart applications of Fast Fourier Transforms to Levin integration, using ordinary and linear differential equations. For an overview of these methods, I refer to \cite{Assassi:2017, Campagne:2017, Schoneberg:2018, Fang:2020, Bella:2021, Leonard:2023} and references therein.

\begin{figure}
  \centering
  \begin{subfigure}[b]{0.49\textwidth}
    \includegraphics[width=\textwidth]{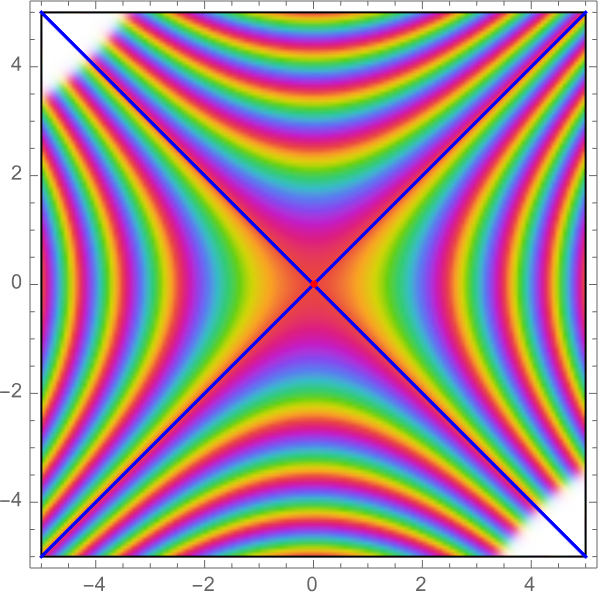}
  \end{subfigure}
  \begin{subfigure}[b]{0.49\textwidth}
    \includegraphics[width=\textwidth]{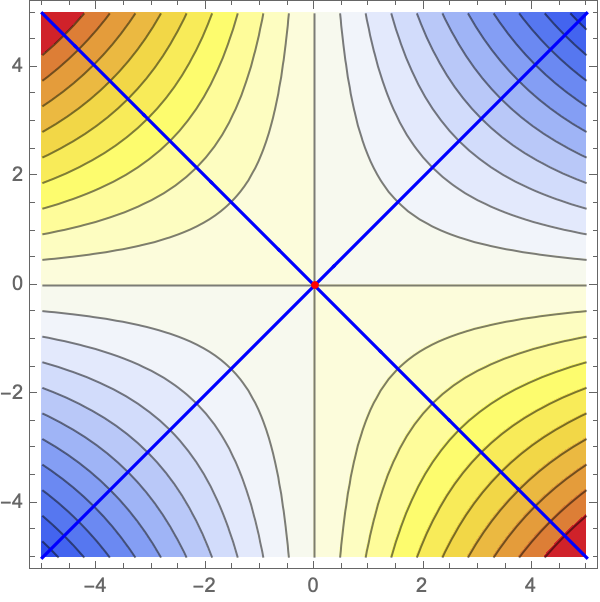}
  \end{subfigure}\\
  \begin{subfigure}[b]{0.49\textwidth}
    \includegraphics[width=\textwidth]{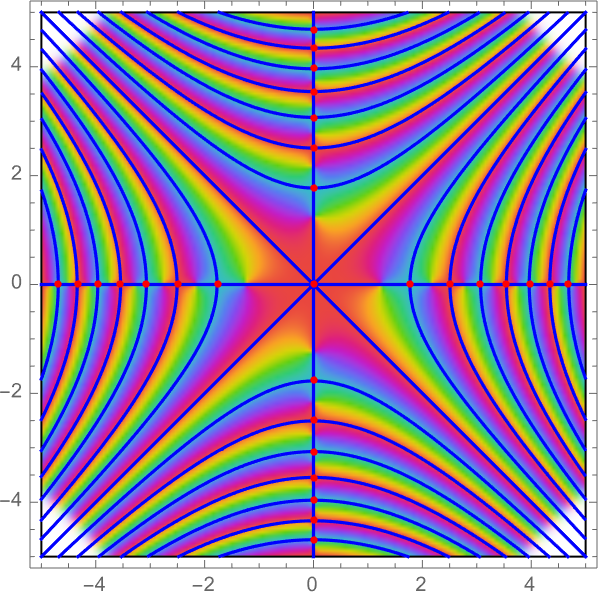}
  \end{subfigure}
  \begin{subfigure}[b]{0.49\textwidth}
      \includegraphics[width=\textwidth]{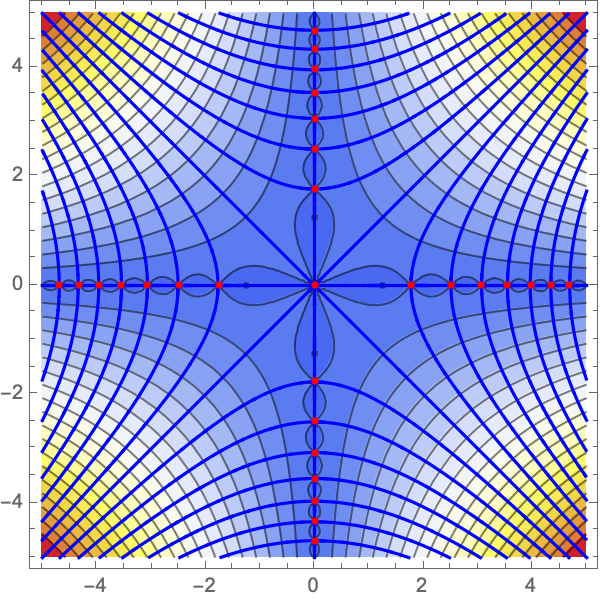}
  \end{subfigure}
  \caption{Picard-Lefschetz theory applied to the integral $\int \exp x^2\mathrm{d}x$ (top) and $\int \cos x^2 \mathrm{d}x$ (bottom) in the complex $x$ plane. The left figures show the saddle points (red points), and the corresponding steepest ascent/descent contours (blue curves) plotted on the complex exponent. The right figures show the same saddle points and the steepest ascent/descent contours plotted on the real part of the exponent, the $h$-function.}
  \label{fig:PL}
\end{figure}

\section{Picard-Lefschetz theory}\label{sec:PicardLefschetz}
Picard-Lefschetz theory is a general method to improve the convergence properties of analytic integrals (first introduced into physics in \cite{Witten:2010}) that makes use of Cauchy's integral theorem. In particular, it formalizes the optimal deformation of oscillatory integrals with analytic integrands yielding the sum of absolutely convergent integrals along a set of integration contours in the complex plane,
\begin{align}
  \int_D e^{ig(x)} \mathrm{d}x = \sum_j \int_{\mathcal{J}_j} e^{ig(x)}\mathrm{d}x\,,
\end{align}
with a sum over the relevant saddle points $x_j$ of $g$ and the steepest descent contours of these saddle points $\mathcal{J}_j$ with respect to the real part of the exponent $h(x)=\text{Re}[ig(x)]$. A saddle point and its associated descent thimble are relevant to the integral if and only if its steepest ascent thimble intersects the origin deformation domain $D$ (assumed the span a region between two singularities of $h$). The integrand $\exp(ig(x))$ does not oscillate along $\mathcal{J}_j$, making the deformation optimal. The deformed integral is easy to evaluate numerically and estimate analytically with the saddle point approximation. For a brief introduction to Picard-Lefschetz theory, see for example \cite{Feldbrugge:2017, Feldbrugge:2023}.

We briefly sketch the Picard-Lefschetz procedure for the Fresnel integral 
\begin{align}
  I = \int_{-\infty}^\infty e^{i x^2}\mathrm{d}x\,.
\end{align}
The integrand has a single relevant saddle point $x_1=0$ with the steepest descent thimble $\mathcal{J}_1=e^{i \pi /4}\mathbb{R}$ and ascent thimble $\mathcal{K}_1=e^{-i \pi /4}\mathbb{R}$ (see the top panels of fig.\ \ref{fig:PL}). As the saddle point is located on the real line, the ascent thimble intersects the real line. Along the thimble, the integral simplifies to the Gaussian integral 
\begin{align}
  I = e^{i \pi /4} \int_{-\infty}^\infty e^{-u^2}\mathrm{d}u = (1+i)\sqrt{\pi/2}\,.
\end{align} 

Note that when applying Picard-Lefschetz theory to the real part of the integrand,
\begin{align}
  \text{Re}[I] = \int_{-\infty}^\infty \cos(x^2)\mathrm{d}x = \sqrt{\pi/2}\,,
\end{align} 
we find that the real line is already the optimal integration domain. The real line is written as an infinite set of steepest descent contours corresponding to the saddle points on the real line $x_n= \pm \sqrt{n \pi}$ (see the bottom panels of fig.\ \ref{fig:PL}), each corresponding to the maxima and minima of the integrand, running between the zero-crossings of $\cos x^2$. At the zero crossings, the real part of the exponent $h$ diverges to $-\infty$. The infinite set of real relevant saddle points \textit{resums} to a single complex saddle point yielding the desired result $\sqrt{\pi/2}$ after taking the real part. 

As we see from this example, the Picard-Lefschetz analysis of a real integral can be dramatically \textit{improved} by introducing an imaginary part to the integrand, \textit{i.e.},
\begin{align}
  \int f(x)\mathrm{d}x = \text{Re}\left[\int(f(x)+ig(x))\mathrm{d}x\right]
  = \text{Re}\left[\sum_j \int_{\mathcal{J}_j}(f(x)+ig(x))\mathrm{d}x\right]\,.
\end{align}
We have the freedom to select a suitable imaginary part $g$. It is generally desirable to formalize the problem with a minimal number of roots of the integrand on the original integration domain, as these generally lead to singularities in the $h$-function and additional saddle points.

\subsection{Gaussian selection kernel}
\begin{figure}
  \centering
  \includegraphics[width=0.5\textwidth]{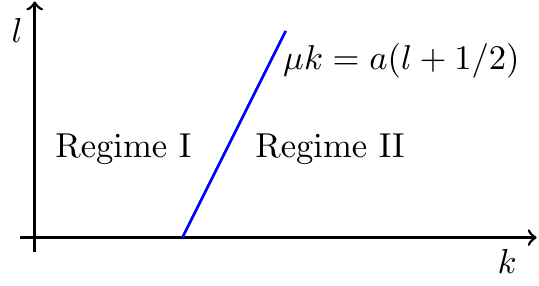}
  \caption{A sketch of the bell-shaped (Regime I) and the oscillatory regime (Regime II) in the $k$-$l$ plane.}
  \label{fig:regimes}
\end{figure}

Now, let's apply these insights to the spherical Bessel transform of the Gaussian selection kernel,
\begin{align}
  F(r) = \frac{1}{\sqrt{2\pi \sigma^2}}e^{-\frac{(r-\mu)^2}{2 \sigma ^2}}\,,
\end{align}
centered at $\mu$ with the standard deviation $\sigma$. In this paper, I will generally assume the kernel to have only significant support for positive proper distances, \textit{i.e.}, $\mu \gg \sigma$. For the spherical Bessel transform
\begin{align}
  \mathcal{F}_l(k) = \int_0^\infty F(r) j_l(k r)\mathrm{d}r\,,
\end{align}
we can distinguish two regimes related to the qualitative behavior of the spherical Bessel function. As we saw in the previous section, the Bessel function $j_l(x)$ vanishes for $x=0$, and slowly rises to reach its first peak at roughly $x=l+1/2$, after which the Bessel function starts to oscillate (see fig.\ \ref{fig:Bessel} for the first few spherical Bessel functions). Consequently, when $ \mu k<a(l + 1/2)$ (\textbf{Regime I}) for an order unity constant $a$, the selection kernel overlaps with the first regime. In this case, the integrand $F(r)j_l(kr)$ is a bell-shaped curve that I evaluate with conventional integration schemes. For $\mu k>a(l + 1/2)$ (\textbf{Regime II}), the selection kernel overlaps with the oscillatory part of the Bessel function. In this paper, I improve the behavior of this integral with the Picard-Lefschetz method. See fig.\ \ref{fig:regimes} for a sketch of the two regimes. In this paper, I use the constant $a = 1$.

\begin{figure}
  \centering
  \begin{subfigure}[b]{0.49\textwidth}
    \includegraphics[width=\textwidth]{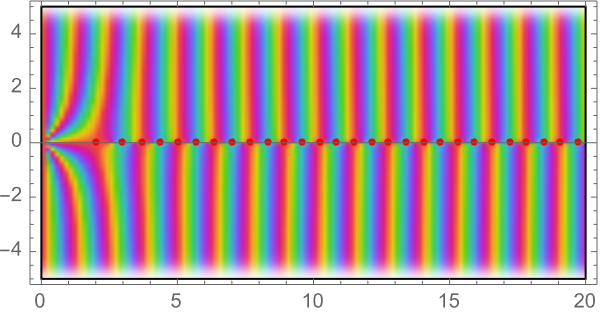}
  \end{subfigure}
  \begin{subfigure}[b]{0.49\textwidth}
    \includegraphics[width=\textwidth]{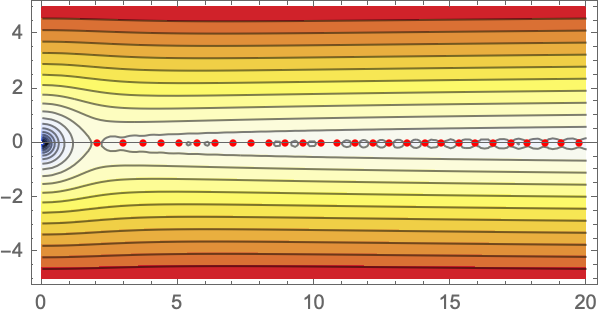}
  \end{subfigure}\\
  \begin{subfigure}[b]{0.49\textwidth}
    \includegraphics[width=\textwidth]{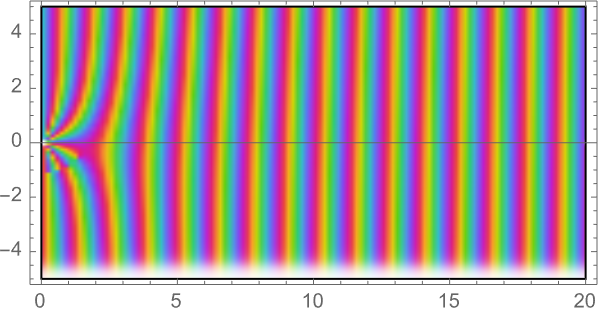}
  \end{subfigure}
  \begin{subfigure}[b]{0.49\textwidth}
    \includegraphics[width=\textwidth]{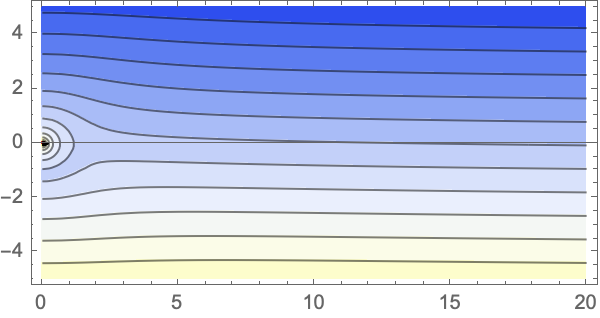}
  \end{subfigure}
  \caption{The analytic continuation of the spherical Bessel function $j_l(kr)$ (top) and the spherical Hankel function of the first kind $h_l^{(1)}(kr)$ (bottom) in the complex $r$ plane, with their saddle points (red points) for $k=5$ and $l=8$. The left panels plot the analytic continuation. The right panels plot the real part of the exponent, the $h$-function.}
  \label{fig:Bessel_Hankel}
\end{figure}

Just like the $\cos x^2$ example discussed in the previous section, the integral over the spherical Bessel function is unaffected by a Picard-Lefschetz deformation. The original integration domain $(0,\infty)$ is already optimal in the Picard-Lefschetz sense. The spherical Bessel function has an infinite number of critical points and zero-crossings. I can simplify the complex structure of the integrand by replacing the spherical Bessel function with the real part of the spherical Hankel function,
\begin{align}
  \mathcal{F}_l(k) = \text{Re}\left[\int_0^\infty F(r) h_l^{(1)}(k r)\mathrm{d}r\right]\,,
\end{align}
where $h_l^{(1)}(x) = j_l(x) + i y_l(x)$ is the spherical Hankel function of the first kind and $y_l$ denotes the spherical Bessel function of the second kind. As we can see in fig.\ \ref{fig:Bessel_Hankel}, the spherical Bessel function has many critical points on the real line. These are absent in the spherical Hankel functions of the first kind. In the limit of large $|x|$, the spherical Hankel function approaches the asymptotic 
\begin{align}
  h_{l}^{(1)}(x) \sim  \frac{e^{i(x- \pi (1+l)/2)}}{x}\,,
\end{align}
which indeed does not vanish for finite $x$ in the complex plane.

\begin{figure}
  \centering
  \begin{subfigure}[b]{0.49\textwidth}
    \includegraphics[width=\textwidth]{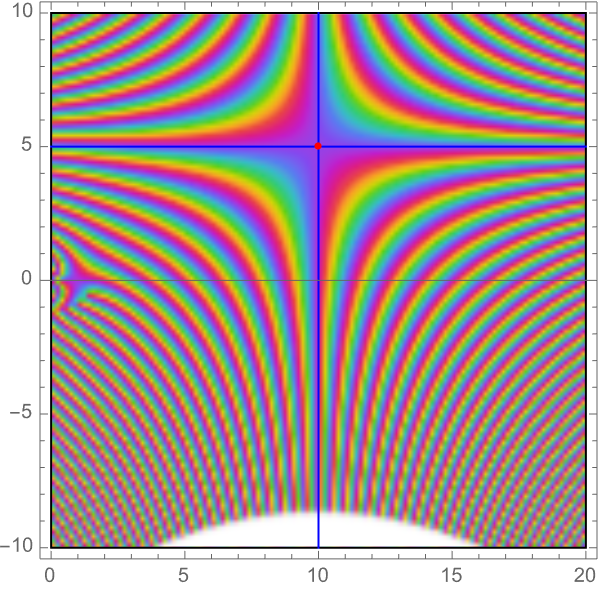}
  \end{subfigure}
  \begin{subfigure}[b]{0.49\textwidth}
    \includegraphics[width=\textwidth]{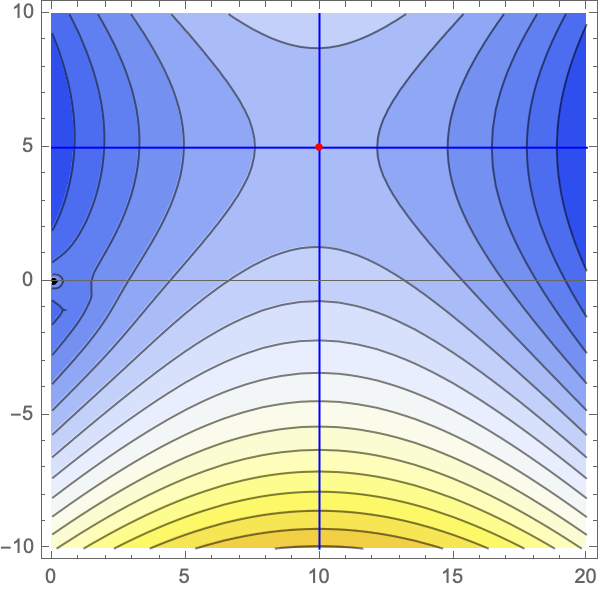}
  \end{subfigure}
  \caption{The analytic continuation of the combination $F(k)j_l(kr)$ for the Gaussian selection kernel with $\mu=10$ and $\sigma = 1$ for $k=5$ and $l=8$ with the saddle point (red point) and the asymptotic steepest descent and ascent contours (blue curves). The left panel plots the analytic continuation. The right panel plots the real part of the exponent, the $h$-function.}
  \label{fig:Gaussian}
\end{figure}

Writing the integrand $F(r)h_l^{(1)}(kr)$ as an exponent $e^{h(r)+i H(r)}$ with the real $h$ and imaginary part $H$, we find that for large $k r$, the $h$-function assumes the asymptotic form
\begin{align}
  h(u+i v) \sim \frac{(v-k \sigma^2)^2 -(u-\mu)^2}{2 \sigma^2} - \frac{\ln (u^2 + v^2)}{2} - \frac{k^2 \sigma^2}{2} - \frac{1}{2}\ln 2 \pi \sigma^2 k^2,
\end{align}
with $r=u+i v$, which is independent of the multipole moment $l$. When ignoring the logarithmic contribution\footnote{Note that the logarithm in $h$ does not appear when considering the selection kernel $F(r)\mapsto rF(r)$. This notwithstanding, I prefer to use the Gaussian selection kernel as it leads to an easier generalization to more general selection kernels using radial bases function interpolation theory.}, the $h$-function has a unique saddle point
\begin{align} 
  r_s \sim \mu + i k \sigma^2\,.
\end{align}
See fig.\ \ref{fig:Gaussian} for an illustration of the structure of the integrand $F(k)j_l(kr)$ in the complex $r$-plane. 
At the saddle point, the integrand assumes the form
\begin{align}
  \mathcal{N} = F(r_s)h_l^{(1)}(k r_s) \sim \frac{1}{\sqrt{2\pi \sigma^2}}\frac{e^{ - \frac{k^2 \sigma^2 }{2}+ i (k \mu- \frac{\pi }{2} l)}}{ik \mu - k^2 \sigma^2}\,.
\end{align}
The first- and second-order derivatives of the exponent approach
\begin{align}
  \alpha = \frac{\partial \ln F(r)h_l^{(1)}(k r)}{\partial r}\bigg|_{r=r_s} &\sim-\frac{1}{\mu+i k \sigma^2}\,, \\
  \beta = \frac{\partial^2 \ln F(r)h_l^{(1)}(k r)}{\partial r^2}\bigg|_{r=r_s} &\sim \frac{1}{(\mu+i k \sigma^2)^2}-\frac{1}{\sigma^2}\,.
\end{align}
The higher-order derivatives decay as 
\begin{align}
  \frac{\partial^n \ln F(r)h_l^{(1)}(k r)}{\partial r^n}\bigg|_{r=r_s} \sim \frac{(-1)^{n+1}n!}{(\mu+i k \sigma^2)^n}\,.
\end{align}
Note that the second-order derivative $|\beta|$ dominates over first- and higher-order derivatives in $r_s$, as $r_s$ is very close to the true saddle point of the exponent.

\begin{figure}
  \centering
  \begin{subfigure}[b]{0.48\textwidth}
    \includegraphics[width=\textwidth]{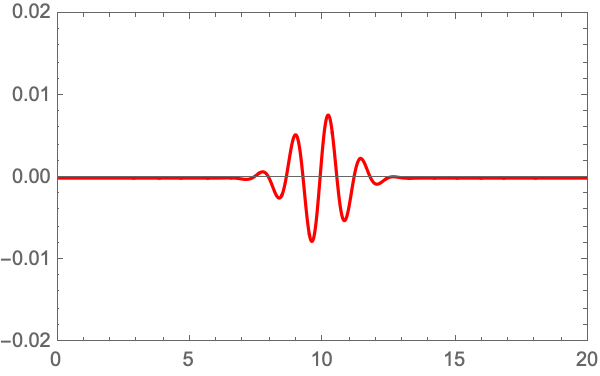}
  \end{subfigure}
  \begin{subfigure}[b]{0.51\textwidth}
    \includegraphics[width=\textwidth]{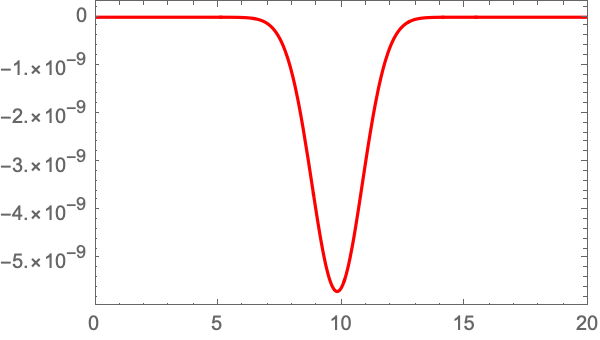}
  \end{subfigure}
  \caption{The real part of the integrand along the original (left) and deformed integration domain (right) for $l=8$, $k=5$, $\mu=10$ and $\sigma=1$.}
  \label{fig:deformed}
\end{figure}

Picard-Lefschetz theory provides the optimal deformation of the oscillatory integral in terms of a set of steepest descent contours. By Cauchy's theorem, the deformation will not alter the integral. However, in practice, the implementation of an integral along the steepest descent contours can be delicate (for a numerical implementation see \url{https://p-lpi.github.io/} and \cite{Feldbrugge:2023}). In this paper I will instead, inspired by Picard-Lefschetz theory, propose an approximation of the descent contour that is easy to implement numerically. Explicitly, I propose to shift the original integration domain $(0,\infty)$ to the contour $(i k \sigma^2, i k \sigma^2 + \infty)$,
\begin{align}
  \mathcal{F}_l(k) = \text{Re}\left[\int_{i k \sigma^2}^{\infty+i k \sigma^2} F(r) h_l^{(1)}(k r)\mathrm{d}r\right]\,,
  \label{eq:complexLimber}
\end{align}
removing most of the oscillations and making the integrand decay exponentially away from the saddle point. See fig.\ \ref{fig:deformed} for an illustration of the integrand evaluated along the real line and the shifted integration domain. This shift is easier to implement than the exact deformation onto the descent contours. Note that we might as well extend the integration domain to $-\infty + i k \sigma^2$ as the integrand is insignificant for negative $r$,
\begin{align}
  \mathcal{F}_l(k) = \text{Re}\left[\int_{-\infty + i k \sigma^2}^{\infty+i k \sigma^2} F(r) h_l^{(1)}(k r)\mathrm{d}r\right]\,.
\end{align}
This equation is exact, assuming the selection kernel is insignificant for negative proper distances, regardless of the asymptotic expansion, as the spherical Hankel function $h_l^{(1)}(x)$ has only a single pole at the origin (corresponding to the divergence of the spherical Bessel function of the second kind $y_l(x)$ in the limit $x \to 0$). This deformed integral can be used to speed up the evaluation of the angular power spectrum $C_{AB}(l)$. In particular, for numerical purposes, the integration domain in $r$ can be restricted to a small interval centered at the approximate saddle point.

It is tempting to claim that equation \eqref{eq:complexLimber} is not only correct when the integrand is oscillatory (\textbf{Regime II}) but also when the integrand follows a bell-shaped curve (\textbf{Regime I}). Although this is formally true, the singularity at the origin and the rapid divergence at the peak of the Gaussian kernel in the complex plane lead to numerical instabilities in \textbf{Regime I}. For this reason, it is preferable to evaluate the spherical Bessel transform directly in \textbf{Regime I} using equation \eqref{eq:SBT} and apply the Picard-Lefschetz definition using equation \eqref{eq:complexLimber} in \textbf{Regime II}. 

The Picard-Lefschetz method can generally be applied to the spherical Bessel transformations of analytic selection kernels. I prefer to restrict the present analysis to the Gaussian kernel, as it leads to a particularly simple deformation of the integration domain. In particular, the same method applies to integrals of the form $\int r^n F(r) j_l(kr)\mathrm{d}r$, as the introduction of a polynomial does not significantly alter the structure of the analytic continuation of the integrand in the complex plane. More general selection kernels can be constructed as a linear combination of these Gaussian features (see the next section and appendix \ref{ap:Radial_interpolation_Theory}). When the selection kernel is the result of numerical computation, the analytic continuation is not available and we will need to resort to interpolation functions anyway. 

The Picard-Lefschetz method is easily extended to integrals involving the $n$-th order derivative of the spherical Bessel function, using the recursion relation
\begin{align}
  \frac{\partial j_l(x)}{\partial x} = -\frac{j_l(x)}{2 x} + \frac{1}{2} \left(j_{l-1}(x) - j_{l+1}(x)\right)\,,
\end{align} 
and the observation that the proposed complex deformation is independent of $l$. This makes the method equally applicable to numerically more challenging cases including the contributions from redshift-space distortions and Doppler effects.

\subsection{General selection kernel}
In the previous section, I studied the spherical Bessel transform of the Gaussian selection kernel. We here extend our study to the linear combination
\begin{align}
  F(r) = \sum_{j=1}^N \omega_j \varphi(r - \mu_j)
\end{align}
of the Gaussian basis function
\begin{align}
  \varphi(x) = \frac{1}{\sqrt{2\pi \sigma^2}}e^{-\frac{x^2}{2\sigma^2}}
\end{align}
centered at $\mu_j$, with the weights $\omega_j$. These curves cover a large space of functions on the real line while keeping tight control of their analytic continuations. In fact, using radial basis function interpolation theory, we can efficiently interpolate a general set of data points $F(\mu_j)=v_j$, with a matrix equation 
\begin{align}
  \bm{\omega}  = M^{-1} \bm{v}\,.
\end{align}
with the weights $\bm{\omega}=(\omega_1,\dots,\omega_N)$, the values $\bm{v}=(v_1,\dots,v_N)$, and the interpolation matrix $M_{i,j}=\varphi(|\mu_i - \mu_j|)$. The interpolation matrix $M$ is invertible when the basis function is \textit{strictly positive definite}. This condition is satisfied by the Gaussian basis function. See appendix \ref{ap:Radial_interpolation_Theory} for a brief sketch of radial basis function interpolation theory.

Using the Gaussian representation of the selection kernel, we can evaluate the spherical Bessel transform using the same deformation of the integration contour,
\begin{align}
  \mathcal{F}_l(k) = \text{Re}\left[\int_{-\infty+i k \sigma^2}^{\infty+i k \sigma^2} \sum_{j=1}^N \omega_j \varphi(r-\mu_j) h_l^{(1)}(k r)\mathrm{d}r\right]\,,
\end{align}
in \textbf{Regime II}. Along the deformed integration domain, the integrand decays exponentially, leading to a quick evaluation with standard numerical methods. In \textbf{Regime I}, the integrand does not oscillate. Here, I evaluate the integral using conventional integration techniques such as the Gaussian quadrature method.

\subsection{Saddle point approximation}
In both \textbf{Regime I} and \textbf{Regime II}, the spherical Bessel transform of the Gaussian selection kernel is expressed in terms of an integral over a bell-shaped integrand. This enables the estimation of the integral with a saddle point approximation. In order to make the approximation more accurate, I refine \textbf{Regime II} into parts.

\begin{figure}
  \centering
  \includegraphics[width=0.48\textwidth]{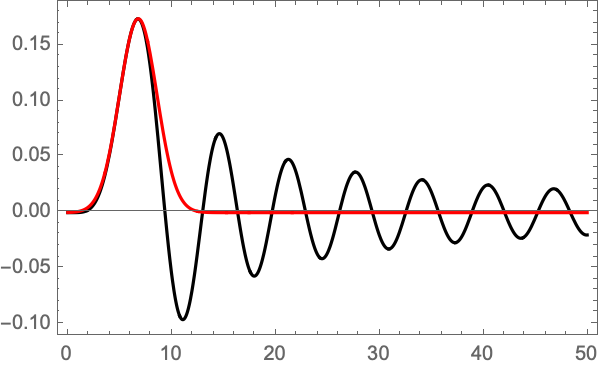}
  \caption{The spherical Bessel function (the black curve) and the Gaussian approximation of the first peak (the red curve) for $l=5$.}
  \label{fig:regimeI}
\end{figure}

\begin{itemize}
  \item \textbf{Regime I} ($\mu k<(l + 1/2)$): I expand the spherical Bessel function $j_l(x)$ around the point $x=l+1/2$ to obtain the Gaussian approximation
  \begin{align}
    j_l(kr) \sim \mathcal{M} e^{\gamma (kr - l-1/2) + \frac{1}{2} \delta(kr - l -1/2)^2}\,,
  \end{align} 
  with $\mathcal{M} = j_l(l+1/2)$, $\gamma = \partial \ln j_l(x)/\partial x|_{x=l+1/2}$ and $\delta = \partial^2 \ln j_l(x)/\partial x^2|_{x=l+1/2}$. This Gaussian approximates the spherical Bessel function up to the first peak (see fig.\ \ref{fig:regimeI}). The spherical Bessel transform of the Gaussian assumes the form
  \begin{align}
    \mathcal{F}_l  \sim \mathcal{M} \int_{-\infty}^\infty F(r) e^{\gamma (kr - l-1/2) + \frac{1}{2} \delta(kr - l -1/2)^2}\mathrm{d}r\,.
  \end{align}

  Alternatively, we can perform a saddle point approximation (like in regime II) at the approximate saddle point $r_s=\mu$. This saddle point approximation becomes increasingly accurate for small $\sigma$.

  \item \textbf{Regime IIa} ($(l + 1/2)\leq \mu k<2(l + 1/2)$): I approximate the spherical Bessel transform with the saddle point approximation
  \begin{align}
    \mathcal{F}_l(k) 
    &\approx 
    \text{Re}\left[\mathcal{N}\int_{-\infty}^\infty e^{\alpha x + \frac{1}{2} \beta x^2}\mathrm{d}x\right]\label{eq:gaussian1}\\
    &=\text{Re}\left[\mathcal{N} \sqrt{\frac{2\pi}{-\beta}} e^{-\frac{\alpha^2}{2\beta}}\right]\label{eq:gaussian2}\,.
  \end{align}
  where $\mathcal{N}$ is defined as the integrand $F(r)j_l(k r)$ and where $\alpha$ and $\beta$ denote the first and second order derivatives of the logarithm of the integrand $\ln F(r)j_l(k r)$ at the approximate saddle point $r_s = \mu$. This is a good approximation for small $\sigma$. Higher-order corrections can be included by expanding the exponent $\ln F(r)j_l(k r)$ further around the point $r_s$. Note that this approximation will fail for large $l$, as the integral will be dominated by an interval around $r=(l+1/2)/k$ (the Limber approximation).

  \item \textbf{Regime IIb} ($2(l + 1/2)\leq \mu k$): I approximate the spherical Bessel transform again with the saddle point approximation \eqref{eq:gaussian1} and \eqref{eq:gaussian2}, where now $\mathcal{N}$ is defined as the integrand $F(r)h_l^{(1)}(k r)$ and $\alpha$ and $\beta$ are the first and second order derivatives of the logarithm of the integrand $\ln F(r)h_l^{(1)}(k r)$ at the approximate saddle point $r_s = \mu + i k \sigma^2$. This is a good approximation for small $\sigma$. Higher-order corrections can be included by expanding the exponent $\ln F(r)h_l^{(1)}(k r)$ further around the saddle point. 
\end{itemize}
Note that this approximation works best for small standard deviation $\sigma$ and small multipole moments where the Limber approximation is not an accurate estimate of the spherical Bessel transform.

For the general selection kernel, built out of Gaussian basis functions, I estimate the spherical Bessel transform as a sum over the approximation of the Gaussian kernels. In \textbf{Regimes IIa} and \textbf{IIb}, the saddle point approximation yields
\begin{align}
  \mathcal{F}_l(k) 
  &\approx \sum_{j=1}^N \omega_j \text{Re}\left[  \mathcal{N}_j \sqrt{\frac{2\pi}{-\beta_j}} e^{-\frac{\alpha_j^2}{2\beta_j}}\right]\,,
\end{align}
with $\mathcal{N}_j$, $\alpha_j$, and $\beta_j$ the integrand and the first and second order derivatives of the logarithm of the integrand $\ln \varphi(r-\mu_j) h_{l}^{(1)}(kr)$ evaluated in the appropriate saddle point depending on the regime. This expression involves one saddle point approximation for every Gaussian basis function.

Just like the Limber approximation, these saddle point approximations reduce the angular power spectrum integral from a three-dimensional integral to a one-dimensional integral over the Fourier mode $k$. The saddle point approximation becomes increasingly accurate with decreasing standard deviation $\sigma$. This leads to the curious proposal of improving the accuracy by decreasing $\sigma$. However, decreasing $\sigma$ generally requires a larger set of basis functions for a fixed selection kernel.

\section{Comparison}\label{sec:Comparison}

\begin{figure}
  \centering
  \begin{subfigure}[b]{0.49\textwidth}
    \includegraphics[width=\textwidth]{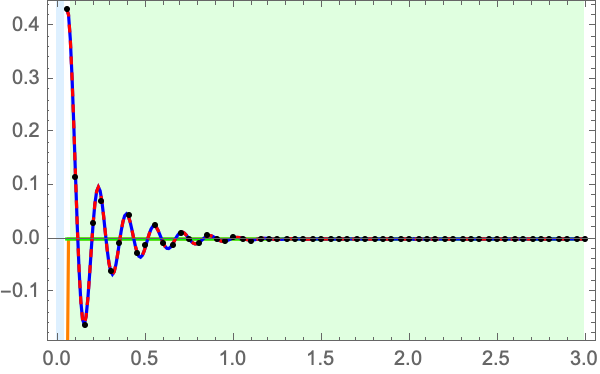}
    \caption{$l=1$}
  \end{subfigure}
  \begin{subfigure}[b]{0.49\textwidth}
    \includegraphics[width=\textwidth]{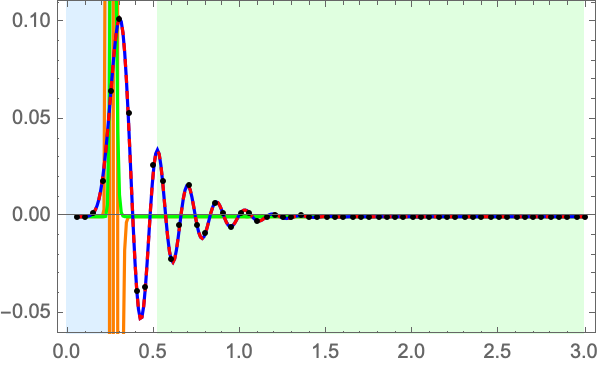}
    \caption{$l=10$}
  \end{subfigure}
  \begin{subfigure}[b]{0.49\textwidth}
    \includegraphics[width=\textwidth]{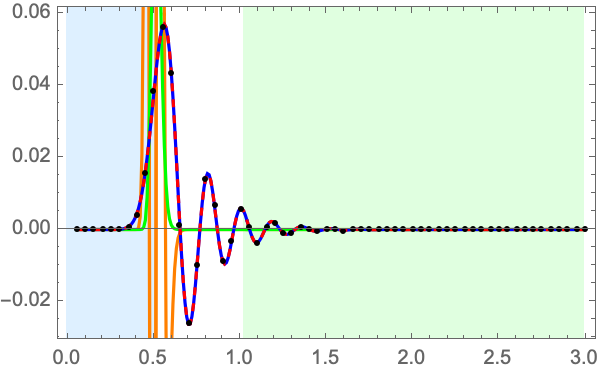}
    \caption{$l=20$}
  \end{subfigure}
  \begin{subfigure}[b]{0.49\textwidth}
    \includegraphics[width=\textwidth]{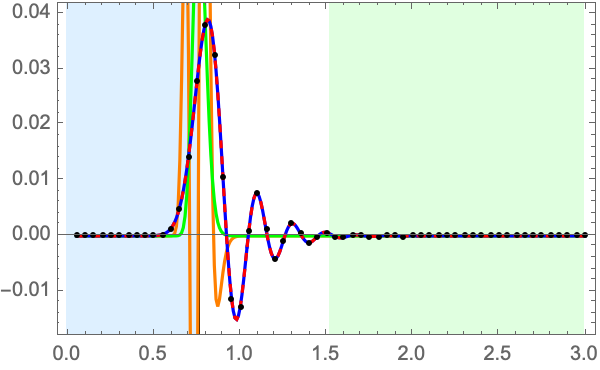}
    \caption{$l=30$}
  \end{subfigure}
  \begin{subfigure}[b]{0.49\textwidth}
    \includegraphics[width=\textwidth]{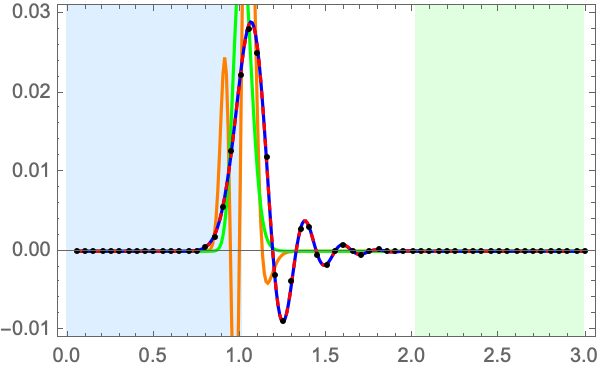}
    \caption{$l=40$}
  \end{subfigure}
  \begin{subfigure}[b]{0.49\textwidth}
    \includegraphics[width=\textwidth]{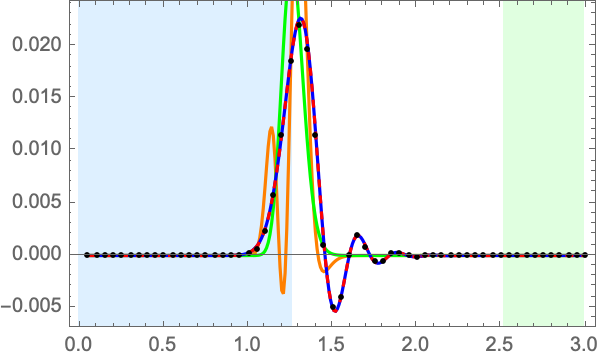}
    \caption{$l=50$}
  \end{subfigure}
  \caption{The spherical Bessel transform for a Gaussian selection kernel with $\mu=40$, $\sigma=2$ for several multipole moments $l$ as a function of the Fourier mode $k$. The exact result (the blue curve), the Limber approximation (the green curve), the extended Limber approximation (the orange curve), the Picard-Lefschetz evaluation (the red curve), and the saddle point approximation (the black points). The blue, white and green regions correspond to \textbf{Regime I}, \textbf{IIa} and \textbf{IIb}.}
  \label{fig:SphericalBesselTransform}
\end{figure}

The accuracy of the proposed Picard-Lefschetz method and the saddle point approximation can be assessed by evaluating the spherical Bessel transform of various Gaussian selection kernels with brute force methods and comparing the result with the Picard-Lefschetz evaluation. In the following, I also compare the proposed saddle point approximation with the Limber approximation and its extension.

In fig.\ \ref{fig:SphericalBesselTransform} we can see the spherical Bessel transform as a function of $k$ for a range of multipole moments $l$. The deformed integral agrees exactly with the original integral in both \textbf{Regimes I} and \textbf{II}. Note that the result peaks near the boundary of the two regimes. The Limber approximation and its extension fail for low multipole moments and approach the true result as the multipole moment increases, converging to the asymptotic
\begin{align}
  \mathcal{F}_l(k)=\sqrt{\frac{1}{4l+2}}\frac{1}{k\sigma} e^{-\frac{(1+2l-2k\mu)^2}{8k^2\sigma^2}}\,.
\end{align}
The saddle point approximation does an excellent job of capturing the behavior of the exact result, both in \textbf{Regimes I} and \textbf{II}.

\begin{figure}
  \centering
  \begin{subfigure}[b]{0.32\textwidth}
    \includegraphics[width=\textwidth]{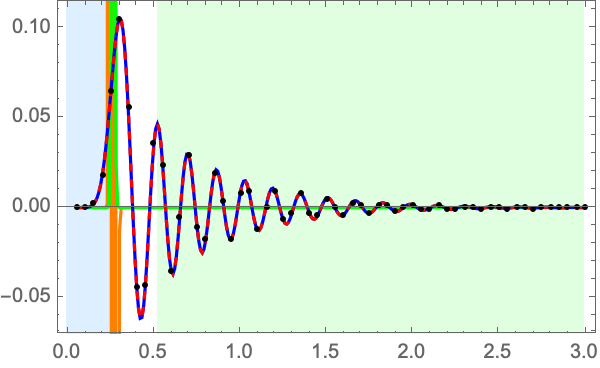}
    \caption{$\sigma=1$}
  \end{subfigure}
  \begin{subfigure}[b]{0.32\textwidth}
    \includegraphics[width=\textwidth]{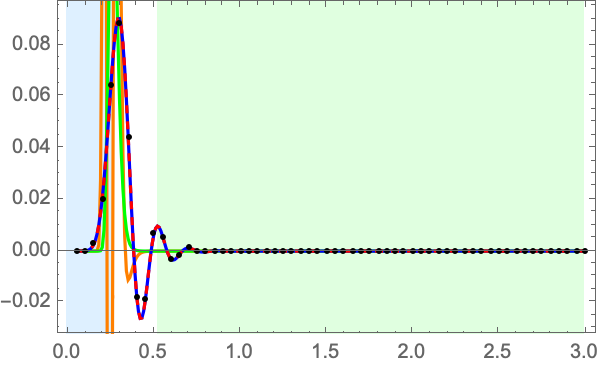}
    \caption{$\sigma=4$}
  \end{subfigure}
  \begin{subfigure}[b]{0.32\textwidth}
    \includegraphics[width=\textwidth]{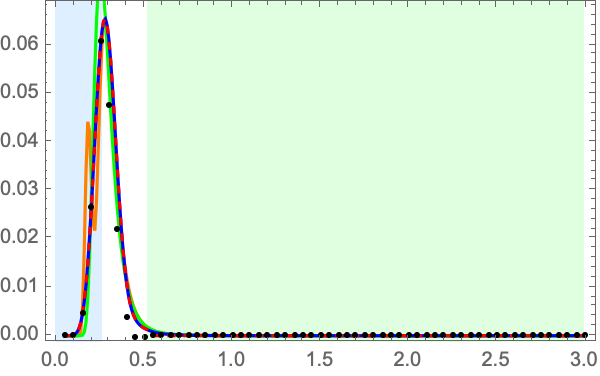}
    \caption{$\sigma=8$}
  \end{subfigure}
  \caption{The spherical Bessel transform for multipole moment $l=10$, and $\mu=40$ for several standard deviations $\sigma$ as a function of the Fourier mode $k$. The exact result (the blue curve), the Limber approximation (the green curve), the extended Limber approximation (the orange curve), the Picard-Lefschetz evaluation (the red curve), and the saddle point approximation (the black points). The blue, white and green regions correspond to \textbf{Regime I}, \textbf{IIa} and \textbf{IIb}.}
  \label{fig:varying_s}
\end{figure}

As we increase the width of the selection kernel, the evaluation of the integral along the deformed integration domain is still accurate, but the saddle point approximation starts to fail (see fig.\ \ref{fig:varying_s}). The fall off of the integrand around the saddle point $r_s=\mu + i k \sigma^2$ is less quick and the integral increasingly receives contributions of the integration domain away from the saddle point. Note that both the Limber approximation and the extended Limber approximation become increasingly accurate in this regime. The saddle point and Limber approximations seem to nicely complement each other. We could improve the accuracy of the saddle point approximation by representing the broader Gaussian as a sum of tighter Gaussian curves using radial basis function interpolation theory.
\bigskip

\begin{figure}
  \centering
  \begin{subfigure}[b]{0.5\textwidth}
    \includegraphics[width=\textwidth]{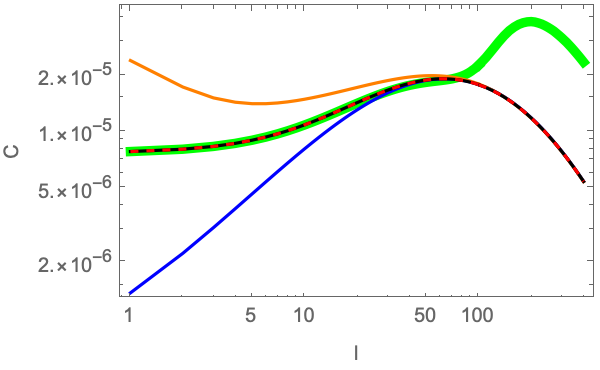}
  \end{subfigure}
  \begin{subfigure}[b]{0.48\textwidth}
    \includegraphics[width=\textwidth]{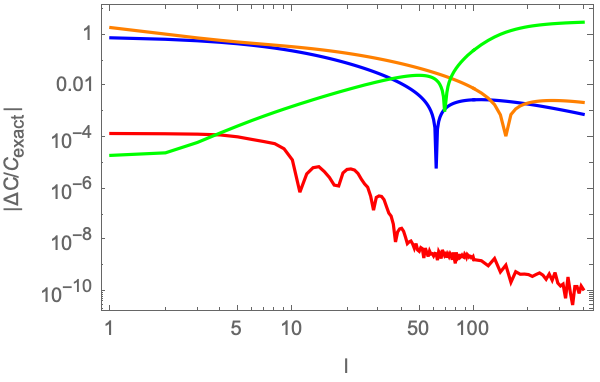}
  \end{subfigure}
  \caption{Angular auto-power spectrum $C(l)$ (left) and the fractional residue $|(C_{exact}(l)-C_{approx}(l))/C_{exact}(l)|$ (right) for a Gaussian selection kernel centered at $\mu=1000$ Mpc with a standard deviation of $\sigma = 50$ Mpc corresponding to a an approximate redshift $z=0.27$ and spread in redshift $\sigma_z=0.05$. I evaluate the angular power spectrum with a brute force evaluation (the black curve), the Picard-Lefschetz evaluation (the red curve, both solid and dashed), the Limber approximation (the blue curve), the extended Limber approximation (the orange curve) and the saddle point approximation (the green curve) of the spherical Bessel transformation.}
  \label{fig:angular}
\end{figure}

The various approximations of the spherical Bessel transform lead to different approximations of the projection of the three-dimensional power spectrum to the angular power spectrum. In fig.\ \ref{fig:angular}, I compare the angular power spectrum corresponding to the Limber approximation, the extended Limber approximation, and the saddle point approximation with the brute force and complex evaluation of the angular power spectrum for a flat $\Lambda$CDM Universe with the fractional matter and dark energy content $\Omega_m = 0.27, \Omega_\Lambda = 0.73$, the current Hubble parameter $H_0 = 67.4$ km/s/Mpc, the scalar fluctuation amplitude $\sigma_8 = 0.8$, and the scalar spectral index $n_s = 0.965$, pushed forward to the linear matter power spectrum using the transfer function of \cite{Eisenstein:1998}.

We clearly see that both the traditional and the extended Limber approximation agree with the brute force evaluation of the angular power spectrum for high multipole moments. The traditional Limber approximation underestimates and the extended Limber approximation overestimates the power at small multipole moments. The saddle point approximation, based on the deformed integral, does a good job of approximating the angular power spectrum for small multipole moments. This approximation fails for high multipole moments. It is striking that the saddle point approximation fails at approximately the same multipole moment where the Limber approximation becomes an accurate approximation of the angular power spectrum. This is related to the observation that the saddle point approximation at the point $r_s=\mu$ in \textbf{Regime IIa} becomes inaccurate for large multi-pole moments when the largest contribution to the integral comes from an interval around $r=(l+1/2)/k$ instead of $r_s=\mu$ (following the Limber approximation). I can improve upon the saddle point approximation in \textbf{Regime IIa} by either improving the approximation of the saddle point or by finding the saddle point numerically. However, at the moment, this seems unnecessary seeing as the Limber approximation is very successful in this particular domain. For smaller $\mu$, the approximate saddle point approximation at $r_s=\mu$ is accurate for larger $l$, bridging the transition to the Limber approximation.

The angular power spectrum evaluated with the complexly deformed spherical Bessel transform agrees very well with the brute force evaluation for all multipole moments. This numerical evaluation is a good alternative to the brute force evaluation when evaluating the angular power spectrum with a three-dimensional integral.

\section{Conclusions}\label{sec:Conclusions}
The spherical Bessel transform is often evaluated in cosmology when projecting the three-dimensional power spectrum onto the angular power spectrum on the celestial sphere. This transformation is generally the most expensive step as it relies on the delicate cancelations of many oscillations. Historically, the Bessel transform was estimated with the Limber approximation that is accurate for large multipole moments \cite{Limber:1953}. More recently, this approximation was extended to include the first derivatives of the selection kernel in the approximation \cite{LoVerde:2008}. Besides these analytic approximations, several numerical schemes for the evaluation of angular power spectra were developed \cite{Assassi:2017, Campagne:2017, Schoneberg:2018, Fang:2020, Bella:2021}.

In this paper, I use Picard-Lefschetz theory to develop an alternative way to evaluate and approximate the spherical Bessel transform of a Gaussian selection kernel in the complex plane. This deformation of the problem, inspired by Picard-Lefschetz theory, resums the oscillatory integral over an infinite set of saddle points into an integral over a single bell-shaped curve. This method works for any multipole moment and leads to an efficient evaluation using conventional numerical integration methods. Inspired by the success of this deformation, I propose a saddle point approximation that works for a large part of the parameter space, complementing the traditional Limber approximation and its extension. This method is especially useful for small multipole moments, narrow redshift bins, or samples with small redshift overlap, for which the Limber approximation has a large error. 

The proposed integration method has yielded a significant improvement over the brute force evaluation of the angular power spectrum. A detailed comparison of the Picard-Lefschetz scheme and the corresponding saddle point method with other proposed schemes for the evaluation of angular power spectra, such as the LogFFT and Levin integration scheme, is beyond the scope of the present paper and will be left for a future investigation. In such a future investigation, I will compare the efficiency of this complex proposal with the schemes, including the ones presented in \cite{Leonard:2023}, and publish an optimized numerical implementation of the complex evaluation of the angular power spectrum in the hope that this will benefit the general cosmological community.

{\bf Acknowledgements:}
I thank Neil Dalal and  Niayesh Afshordi for raising my interest in the numerical evaluation of angular power spectra. I thank Dylan Jow, Ue-Li Pen and Neil Turok for our discussions on Picard-Lefschetz theory and oscillatory integrals in general. The work of JF is supported by the STFC Consolidated Grant `Particle Physics at the Higgs Centre,' and, respectively, by a Higgs Fellowship and the Higgs Chair of Theoretical Physics at the University of Edinburgh.

For the purpose of open access, the author has applied a Creative Commons Attribution (CC BY) license to any Author Accepted Manuscript version arising from this submission.

\bibliographystyle{utphys}
\bibliography{Library}

\appendix
\section{Radial basis interpolation function theory}\label{ap:Radial_interpolation_Theory}
\begin{figure}
  \centering
  \begin{subfigure}[b]{0.49\textwidth}
    \includegraphics[width=\textwidth]{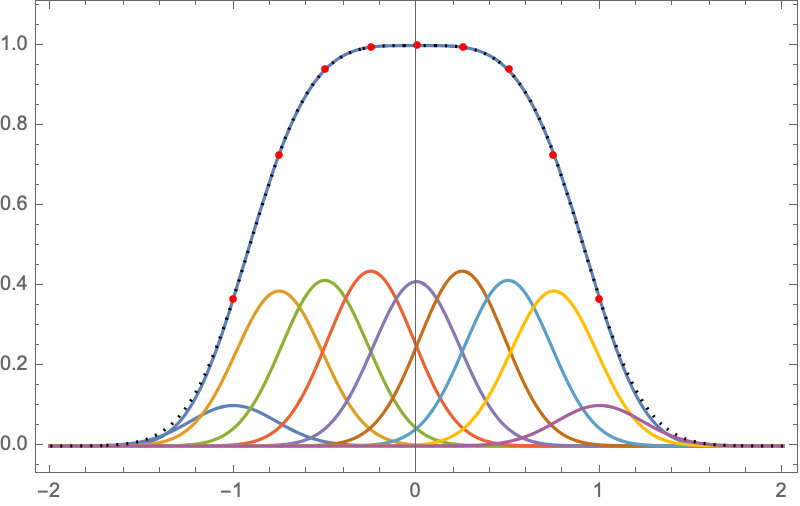}
  \end{subfigure}
  \begin{subfigure}[b]{0.49\textwidth}
    \includegraphics[width=\textwidth]{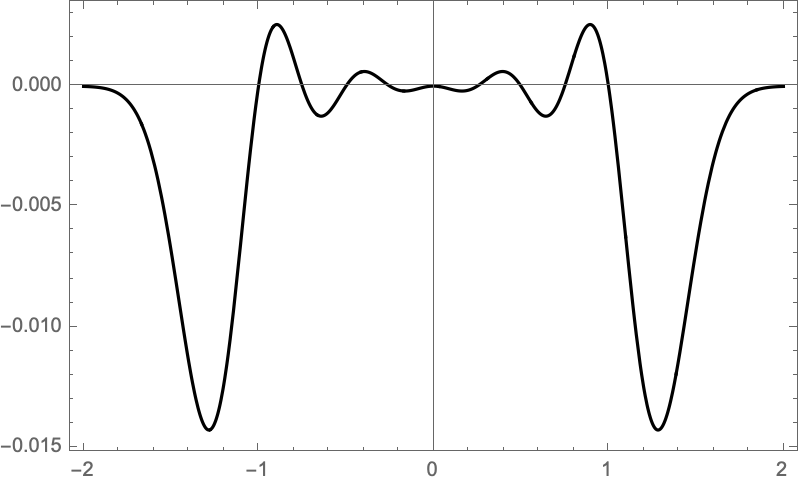}
  \end{subfigure}
  \caption{\textit{Left:} The radial interpolation (black dashed curve) of the function $\exp(-r^4)$ (the blue curve) evaluated at the nine points (the red points) by nine equally spaced Gaussians $\exp(-9 r^2)$ (from left to right, the blue till the purple curves). \textit{Right:} The difference between the function and its radial interpolation  $\exp(-r^4)- f(r)$ as a function of $r$.}
  \label{fig:Radial_Interpolation}
\end{figure}

Radial basis interpolation is an interpolate method where the interpolation function is a linear combination of basis functions (first developed by \cite{Hardy:1971}). Given a set of points $\{r_i\}_{i=1}^N$ and an associated set of function values $f(r_i)= v_i$, we can construct the radial basis interpolation function
\begin{align}
  f(r) = \sum_{j=1}^N \omega_j \varphi(|r - r_j|)\,,
\end{align}
where the weights $\bm{\omega}=(\omega_1,\dots,\omega_N)$ satisfy the matrix equation
\begin{align}
  M \bm{\omega} =  \bm{v}\,,
\end{align}
with the vector $\bm{v}=(v_1,\dots, v_N)$ and the interpolation matrix
\begin{align}
 M =
  \begin{pmatrix}
    \varphi(|r_1 - r_1|) & \varphi(|r_2 - r_1|) & \dots & \varphi(|r_N - r_1|)\\
    \varphi(|r_1 - r_2|) & \varphi(|r_2 - r_2|) & \dots & \varphi(|r_N - r_2|)\\
    \vdots & \vdots & \ddots & \vdots \\
    \varphi(|r_1 - r_N|) & \varphi(|r_2 - r_N|) & \dots & \varphi(|r_N - r_N|)\\
  \end{pmatrix}.
\end{align}
The interpolation matrix is invertible when the basis function $\varphi(x)$ is a \textit{strictly positive definite function}\footnote{A function $f:\mathbb{R} \to \mathbb{C}$ is strictly positive definite when the matrix with components $A_{ij} = f(r_i - r_j)$ is strictly positive definite for any set or real numbers $r_1,\dots,r_N$. In practice, we can often use Bochner's theorem to asses whether a basis is strictly positive definite using its Fourier transform.} \cite{Chang:1996}. Examples of strictly positive definite basis functions are the Gaussian function
\begin{align}
  \varphi(r) = e^{- (\epsilon r)^2}\,,
\end{align}
the Lorentzian function
\begin{align}
\varphi(r) = \frac{1}{1+r^2}\,,
\end{align}
the multiquadric function
\begin{align}
  \varphi(r) = \sqrt{1+ (\epsilon r)^2}\,,
\end{align}
and the bump function
\begin{align}
  \varphi(r) = 
  \begin{cases}
    \exp\left[\frac{-1}{1-(\epsilon r)^2}\right] & \text{for } r < 1/\epsilon\,,\\
    0 & \text{otherwise}\,,
  \end{cases}
\end{align}
with compact support. For strictly positive definite basis functions, the weights can efficiently be evaluated with the matrix equation
\begin{align}
  \bm{\omega}  = M^{-1} \bm{v}\,.
\end{align}
See fig.\ \ref{fig:Radial_Interpolation} for an illustration, approximating the function $\exp(-r^4)$, evaluated on a regularly spaced lattice, with a sum of regularly spaced Gaussian basis functions.
\end{document}